\documentclass[aps,prb,showpacs,superscriptaddress,twocolumn]{revtex4}
\usepackage{amsmath}
\usepackage{amsfonts}
\usepackage{graphicx}
\usepackage{psfrag}

\begin{document}

\title{Stability of spintronic devices based on quantum ring networks}
\author{P\'eter F\"oldi}
\email{foldi@physx.u-szeged.hu}
\affiliation{Department of Theoretical Physics, University of Szeged, Tisza Lajos k\"{o}r%
\'{u}t 84, H-6720 Szeged, Hungary}
\author{Orsolya K\'alm\'an}
\affiliation{Department of Quantum Optics and Quantum Information, Research Institute for Solid
State Physics and Optics,Hungarian Academy of Sciences, Konkoly-Thege Mikl%
\'{o}s \'{u}t 29-33, H-1121 Budapest, Hungary}
\author{F.~M.~Peeters}
\affiliation{Departement Fysica, Universiteit Antwerpen, Groenenborgerlaan 171, B-2020
Antwerpen, Belgium}

\begin{abstract}
Transport properties in mesoscopic networks are investigated, where the strength of the (Rashba-type) spin-orbit coupling is assumed to be tuned with external gate voltages. We analyze in detail to what extent the ideal behavior and functionality of some promising network-based devices are modified by random (spin-dependent) scattering events and by thermal fluctuations. It is found that although the functionality of these devices is obviously based on the quantum coherence of the transmitted electrons, there is a certain stability: moderate level of errors can be tolerated. For mesoscopic networks made of typical semiconductor materials, even cryogenic temperatures can smear out the desired transport properties. When the energy distribution of the input carriers is narrow enough, it turns out that the devices can operate close to their ideal limits even at relative high temperature. As an example, we present results for two different networks: one that realizes a Stern-Gerlach device and another that simulates a spin quantum walker. Finally we propose a simple network that can act as a narrow band energy filter even in the presence of random scatterers.
\end{abstract}

\maketitle

\section{Introduction}
Mesoscopic networks with Rashba-type\cite{R60} spin-orbit coupling have recently gained considerable attention due to their remarkable spin transformation properties. Theoretical proposals for using these networks\cite{KFBP08c,KKF09,SMC09,AWT08,KFBP08b,ZW07,WXE06,MVP05} as specific spintronic devices often focus on the ballistic regime at low (zero) temperatures. Obviously, important features as well as fundamental aspects of the spin-dependent transport problem can be found using this approach, but it is also desirable to investigate the stability of the results in a more realistic context. In this paper we focus on some promising network-based spintronic devices and study how random scatterers  modify the ballistic transport and investigate the extent to which the functionality of these devices remains close to the ideal one. In this aspect, the case when the input electrons are in thermal equilibrium at a finite temperature, i.e., they are not monoenergetic, is also investigated.

The concrete model systems we have chosen to analyze consist of quantum rings that are made of semiconducting materials exhibiting Rashba-type\cite{R60} spin-orbit interaction\cite{G00,NATE97,SKY01,KTHS06} (SOI).
The strength of the SOI that determines the spin sensitive behavior in these rings can be tuned with external gate voltages\cite{NATE97}, offering possible spintronic applications already with two- or three-terminal single quantum rings\cite
{BIA84,X92,NMT99,YPS03,FHR01,MPV04,FR04,FMBP05,WV05,SzP05,SN05,KMGA05,SP05,VKN06,BO07,VKPB07,CHR07,BO08,WC08,CPC08}.
Systems of quantum rings can operate as multi-purpose, flexible spintronic devices\cite{KFBP08c}, and the concept of quantum walk (QW) that inherently requires large systems can also be realized in appropriate ring networks\cite{KKF09}. In this paper we focus on these latter proposals, investigate the stability of rectangular networks of quantum rings as well as that of the networks that can realize QW with the spin degree of freedom playing the role of the quantum coin.

The present paper is organized as follows. In Sec.~II we summarize the theoretical network model to be used, and describe the method for taking scattering events and finite temperatures into account. The geometry and possible functionality of the networks to be investigated is described briefly in Sec.~III. Questions related to the stability of these arrays are investigated in detail in Secs.~IV and V. We summarize our results in Sec.~VI.

\section{Model}
First we consider a single ring \cite{ALG93} of radius $a$ in the $x-y$ plane and assume
a tunable electric field in the $z$ direction controlling the
strength of the spin-orbit interaction characterized by the parameter $%
\alpha $ \cite{NATE97}. The Hamiltonian \cite{Meijer,MPV} in the presence of
spin-orbit interaction for a charged particle of effective mass $m^{\ast}$
is given by
\begin{equation}
H=\hbar\Omega\left[ \left( -i\frac{\partial}{\partial\varphi}+\frac {%
\omega}{2\Omega}\sigma_{r}\right) ^{2}-\frac{\omega^{2}}{4\Omega^{2}}%
\right],  \label{Ham}
\end{equation}
where $\varphi$ is the azimuthal angle of a point on the ring and the radial spin operator is given by $\sigma
_{r}/2=(\sigma_{x}\cos\varphi+\sigma_{y}\sin\varphi)/2.$  The frequency associated with the
spin-orbit interaction is denoted by $\omega=\alpha /\hbar a$, while $\hbar\Omega =\hbar^{2} /2m^{\ast }a^{2}$ gives
the kinetic energy with \thinspace $m^{\ast }$ being the effective mass of
the electron. The position dependent eigenspinors and the corresponding energies of this Hamiltonian can be determined analytically, see e.g.~Refs.~[\onlinecite{MPV,KFBP08}] for details.

When there are leads attached to the ring that connect it to electric contacts, the usual question is the conductance as a function of the SOI strength and/or the energy of the incoming electrons. At zero temperature this latter has zero variance, it equals the Fermi energy $E_F$ of the system. Therefore energy conservation requires finding eigenevalues of the Hamiltonian (\ref{Ham}) being equal to $E_F.$ As a consequence of the two possible eigenspinor orientations as well as the possibility of the current flowing clockwise and counterclockwise along the ring, $E_F$ is usually fourfold degenerate. In order to complete the solution of the scattering problem at a given energy, the spinor valued wave functions have to be joined together. The spinor components have to be continuous, and the spin current \cite{MPV} that enters a junction, should also leave it, i.e., the net current density at a certain junction has to vanish. Let us note that besides Griffith's boundary conditions\cite{G53} that have been described above (and will be used throughout this paper) there are other physically realistic and often used possibilities as well. The choice of the boundary conditions is in fact shown to be related to the reduction of a two dimensional problem to one dimension \cite{V09}.

The procedure above is completely analytic, the transmission matrix $\mathcal{T}$ that connects the input and output spinors as well as the conductance using the Landauer-B\"{u}ttiker formula\cite{D95} can be given in closed forms. This method can be transferred to systems consisting of several rings in a straightforward way. A general scheme for networks to be investigated  can be mathematically represented as
\begin{equation}
|\psi_{in}\rangle\rightarrow \left\{
\begin{aligned}
&|\psi_{out}^1\rangle=\mathcal{T}_1|\psi_{in}\rangle, \\
&|\psi_{out}^2\rangle=\mathcal{T}_2|\psi_{in}\rangle, \\
&\vdots\\
&|\psi_{out}^N\rangle=\mathcal{T}_N|\psi_{in}\rangle,
\label{scheme}
\end{aligned}
\right.
\end{equation}
thus the input state is transformed into $N$ output spinors (and a reflected one, not shown above). In general the transmission matrices $\mathcal{T}_1,\ldots, \mathcal{T}_N$ are different for different outputs\cite{KFBP08b} (see Figs.~\ref{NNfig} and \ref{walkfig} for concrete geometries). Considering the size of these networks, it turns out to be practical to solve the related systems of linear equations numerically. In this way one can determine the spin transformation properties of networks of considerable size, and may find geometries with possible spintronic applications.

\bigskip

However, in a large, or even mesoscopic system the transport process will not be necessarily ballistic, the quantum mechanical coherence of the carrier wave functions may not be maintained all along the device. In order to give account for this issue, we introduce independent point-like scattering centers in the rings. That is, we assume the presence of an additional potential
\begin{equation}
U_{scatt}^{(1)}(\varphi)=\sum_n U_n(D) \delta(\varphi-\varphi_n), \label{Scattpot1}
\end{equation}
with uniformly distributed independent random positions $\varphi_n$ of the Dirac delta peaks. For the sake of simplicity, the density of the scatterers is chosen such that there is always a single one between two neighboring junctions, that is, e.g.~for a four terminal ring we have four scatterers. The strengths of the potentials, $U_n(D),$ are random, they are drawn from independent normal distributions, with zero mean and root-mean-square deviation $D.$ That is, the probability for $U_n(D)$ to have a value in a small interval around $u$ is given by $p(u)du$, where
\begin{equation}
p(u)=\exp(-u^2/2D^2)/D\sqrt{2\pi}. \label{normal}
\end{equation}
In this way, by tuning $D$ we can model weak disturbances (small $D$) as well as the case when frequent scattering events completely change the character of the transport process (corresponding to large values of $D$).

As the interference phenomena that are responsible for the desired behaviors of the devices are spin sensitive, we also introduce spin-dependent (SD) scatterers
\begin{equation}
U_{scatt}^{(2)}(\varphi)=\sum_n \boldsymbol{U}_n(D) \delta(\varphi-\varphi_n), \label{Scattpot2}
\end{equation}
where $\boldsymbol{U}_n(D)$ now represents a $2\times 2$ diagonal matrix, with independent random diagonal elements $U_{n1}(D)$ and $U_{n2}(D)$.

In both cases, we use the same boundary conditions as at the junctions, thus we require the continuity of the spinor valued wave functions $\Psi_L(\varphi),$ $\Psi_R(\varphi)$ at the left and right side of the scatterer situated at $\varphi_n$. Then the requirement of vanishing spin current density reads:
\begin{equation}
\left.\frac{\partial}{\partial\varphi}\Psi_L(\varphi)\right\vert_{\varphi=\varphi_n}
-\left.\frac{\partial}{\partial\varphi}\Psi_R(\varphi)\right\vert_{\varphi=\varphi_n}=
\left. \frac{\boldsymbol{U}_n(D)}{\hbar\Omega}\Psi_R(\varphi)\right\vert_{\varphi=\varphi_n}. \label{deltafit}
\end{equation}
These additional equations are still linear ones, thus we can find analytical solutions for any random sets $\{\boldsymbol{U}_n(D)\},$ $\{\varphi_n\}.$ That is, for any input state we can calculate the output spinor, that we write symbolically as $\left|\Psi_{out}\{\boldsymbol{U}_n(D),\varphi_n\}\right\rangle.$ When after $M_c$ computational runs, the estimated density operator
\begin{equation}
\begin{aligned}
&\rho_{out}(D)=\\
&\frac{1}{M_c} \sum  \left|\Psi_{out}\{\boldsymbol{U}_n(D),\varphi_n\}\right\rangle
\left\langle\Psi_{out}\{\boldsymbol{U}_n (D),\varphi_n\}\right| \label{rhoscatt},
\end{aligned}
\end{equation}
converges, we have all the possible information needed to describe what effects result from the disturbances characterized by the variable $D.$ (Clearly, the sum above runs over different positions $\{\varphi_n\}$ and scatterer strengths $\{\boldsymbol{U}_n\}$, so that the independent random variables in the latter case correspond to normal distributions with the same root-mean-square deviation $D$.) Note that $\rho_{out}(D)$ is not normalized, we can consider it as a conditional density operator that describes the state of the electron if it is transmitted at all. Using the transmission probability $T,$ we have $\mathrm{Tr} [\rho_{out}(D)] /T=1.$

In the description of the process above it was implicitly assumed that the incoming electrons are monoenergetic, but usually, at finite temperatures, this is not the case. Therefore, if we would like to take the energy distribution of the input electrons also into account, we have to average over all possible input energies. In thermal equilibrium at temperature $T$, the output density operator can be written as
\begin{equation}
\rho_{out}(T)=\int p(E,T) \left|\Psi_{out}(E)\right\rangle
\left\langle\Psi_{out}(E) \right| dE, \label{rhoT}
\end{equation}
where $p(E,T)=-\frac{\partial}{\partial E}[1+\exp{(E-E_F)/k_B T}]^{-1}.$ (Note that this expression corresponds to the Landauer-B\"{u}ttiker formula for the conductance at finite temperature and low bias\cite{D95}.) In practice, we convert the integral (\ref{rhoT}) to a sum over the possible energies, thus the expression for $\rho_{out}(T)$ is similar to Eq.~(\ref{rhoscatt}), but the weights of the projectors are not uniform, they are determined by the Fermi distribution.

\section{Networks and their possible applications}
\begin{figure}[tbh]
\includegraphics[width=8cm]{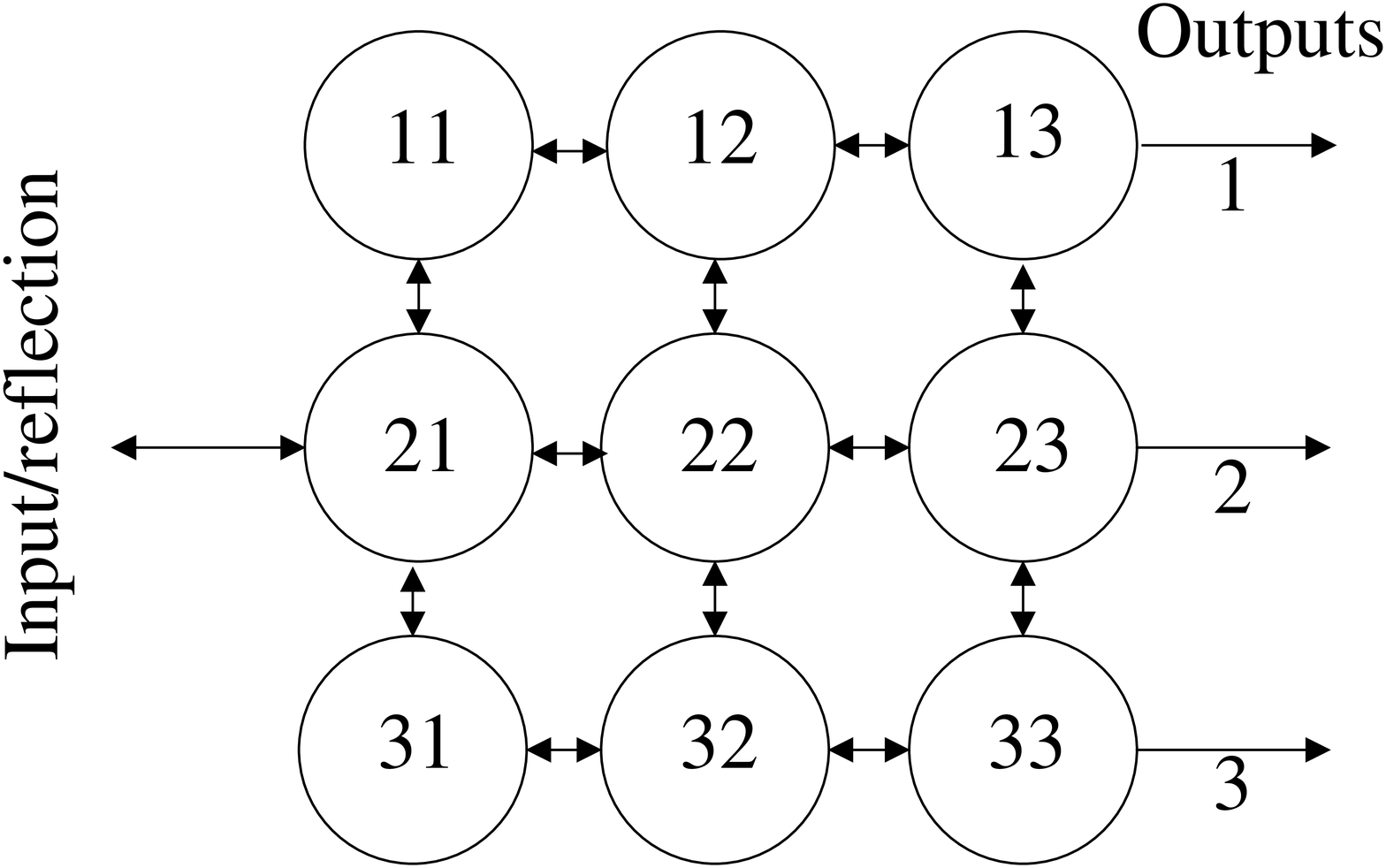}
\caption{The geometry of a rectangular ring network. The arrows indicate the possible directions of
the currents.}
\label{NNfig}
\end{figure}

In this section we summarize briefly the properties of the networks the stability of which will be analyzed later.
The first network can be seen in Fig.~\ref{NNfig}, it is a rectangular array of $N\times N$ quantum rings, with one input lead and $N$ outputs. This geometry has already been realized experimentally\cite{BKSN06}, and the conductance properties have also been investigated in theoretical works\cite{ZW07,KFBP08b}. It has been shown in Ref.~[\onlinecite{KFBP08c}] that if the strength of the SOI can be modulated locally (ring by ring), then relatively small ($N=3, N=5$) networks are remarkably versatile from the viewpoint of spin transformations. Working in a given network geometry, the input current can be directed
to any of the output ports, simply by changing the SOI strengths. This kind of operation can already be achieved by a $3 \times 3$ geometry in a practically reflectionless way, and the probability for an electron to leave the device through a lead other than the distinguished one is less than 1\%. A slightly larger network ($N=5$) can also be made completely analogous to the Stern-Gerlach device: If the input is one of the eigenstates of $\sigma_z$ (e.g.~spin-up in this direction) the output will have the same spin direction at a certain output port (with probability higher than 99\%). When spin direction of the input is the opposite (e.g.~spin-down), it is directed towards a different output port, with its final spin direction being the same as the initial one\cite{KFBP08c}.
\bigskip

Having chosen a certain $N\times N$ geometry, one may want to realize a given spin transformation (function of the device). In order to achieve the desired way of operation, there are $N^2$ parameters to adjust, namely the SOI strengths in the rings. At this point, it is reasonable to define a real valued (preferably experimentally measurable) function over this $N^2$ dimensional parameter space that has (a global) minimum when the desired goal is achieved. Using this function, an appropriate multidimensional minimum searching algorithm can find the relevant parameter values. (Note that in principle this procedure can also be applied in experimental situations.) Clearly, the key point here is the choice of the function to be minimized. It was found that rather straightforward definitions work properly with a "downhill simplex" multidimensional minimizing routine\cite{NRec}. If the purpose is to direct the input current into output port $k$, irrespectively from the spin state, then
\begin{equation}
f_1=2-P_k(\mathrm{in:}\uparrow)-P_k(\mathrm{in:}\downarrow)
\end{equation}
is an appropriate choice for the function to be minimized.
Here $P_k(\mathrm{in:}\uparrow)$ [$P_k(\mathrm{in:}\downarrow)$] is the probability of transmission into the distinguished output port, provided the input was spin up (down) in the $z$ direction. That is, $f_1$ is zero if a spin up input wave leaves the network through the distinguished output with 100\% probability {\em and} the same holds for the orthogonal (spin-down) input. (Note that considering the two different inputs, two computational/experimental runs are needed to obtain $f_1.$) It is also worth noting that $f_1$ depends on the SOI strengths in a complex way, via the transmission probabilities, which can be measured.

When our aim is a device that can be the spintronic analogue of the Stern-Gerlach apparatus, we can use the function
\begin{equation}
f_2=2-P_1(\mathrm{in:}\uparrow, \mathrm{out:}\uparrow)-P_5(\mathrm{in:}\downarrow, \mathrm{out:}\downarrow),
\end{equation}
where $P_1(\mathrm{in:}\uparrow, \mathrm{out:}\uparrow)$ [$P_5(\mathrm{in:}\downarrow, \mathrm{out:}\downarrow)$] denotes the transmission probability of a spin up (down) state through output port 1 (5), provided the input was spin up (down). Note that the theoretical minimum of both $f_1$ and $f_2$ is zero and values below $0.05$ can be reached with SOI strengths in the experimentally achievable range.

\begin{figure}[tbh]
\includegraphics[width=8cm]{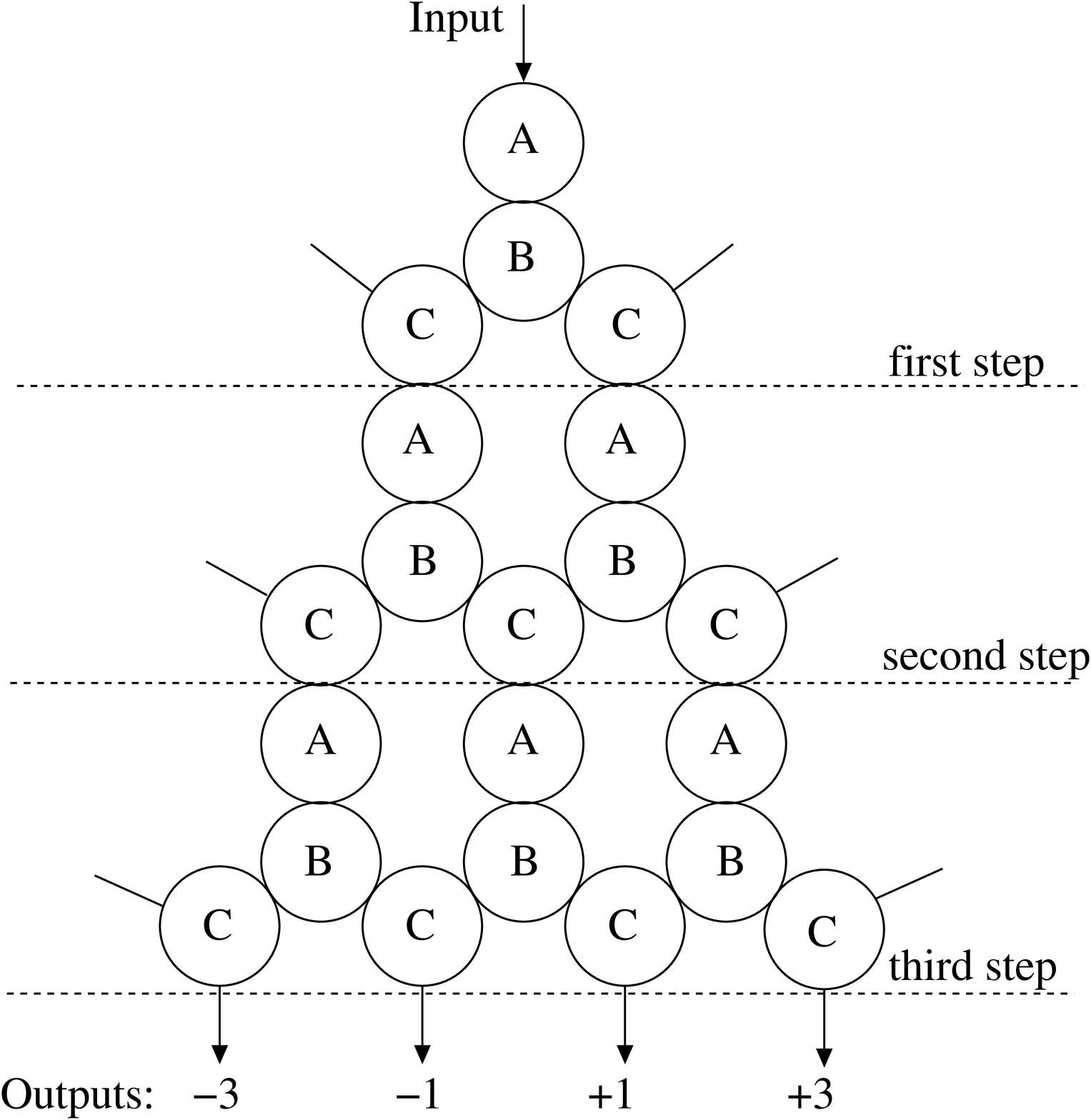}
\caption{Schematic representation of a device that can realize coined quantum walk on the line. The example shown in the figure corresponds to a QW with $M=3$ steps. Note that in the ideal case the current flows in the network in the downward direction, and the input current is distributed among the numbered output ports.}
\label{walkfig}
\end{figure}
\bigskip
The second network geometry we will analyze is shown in Fig.~\ref{walkfig}. The aim of this device is the realization of quantum walk (QW), so that spin degree of freedom plays the role of the quantum coin, i.e., it determines the direction of the steps taken by the walker. The scheme itself is similar to the classical random walk on the line, when the walker tosses a coin and takes a step to the left if the coin is heads or to the right if the coin is tails (or vice versa). In the quantum analogue of such a walk a quantum coin (electron spin, in our case) is used, the state
of which can be a linear combination of the classical heads and tails, or mathematically, any state of a 'coin' Hilbert
space $\mathcal{H}_{C}$, spanned by the two basis states $ \left| L \right>$ and $\left| R \right> $, where $L$ ($R$) stand for 'left' ('right'). The positions of the walker also span a Hilbert space $\mathcal{H}_{P}=\left\lbrace \left| i \right> : i \in \mathbf{Z} \right\rbrace $ with $\left| i \right>$ corresponding to the walker localized in position $i$. The states of the total system are in the space $\mathcal{H}=\mathcal{H}_{C} \otimes\mathcal{H}_{P},$ which can be identified with spinor valued wave functions in our case. The conditional step of the walker depends on the state of the coin, it can be described by the unitary operation
\begin{equation}
 S = \left| L \right> \left< L \right| \otimes \sum_{i} \left| i+1 \right>
 \left< i \right| + \left| R \right> \left< R \right| \otimes \sum_{i}
 \left| i-1 \right> \left< i \right|. \label{S}
\end{equation}
The coin-toss is realized by a unitary operation $C$ acting on the spin states. The QW of $N$ steps is
defined as the transformation $U^{N}$, where $U$, acting on $\mathcal{H}=\mathcal{H}_{C} \otimes \mathcal{H}_{P}$ is given by
\begin{equation}
 U = S \cdot \left( C \otimes I \right). \label{QWstep}
\end{equation}
The network shown in Fig.~\ref{walkfig} realizes this scheme in the following manner: first A-type rings perform Hadamard operation on the spin states. (In other words, the operator $C$ in this scheme is a frequently used, so called balanced unitary coin, which is represented by a matrix in which each element is of equal magnitude.) With appropriately chosen parameters, there are two input spin states $\left| L \right>$ and $ \left| R \right>$ for B-type rings so that $\left| L \right> (\left| R \right>)$ has zero probability to be transmitted into the right (left) output arm. C-type rings have double role in the scheme, first they undo the rotation performed by the B-type rings (B-type rings direct the input states $\left| L \right>$ and $\left| R \right>$ into their appropriate output, but meanwhile they also rotate them) as well as allowing for the interference of subsequent steps by accepting two inputs. Using these three types of rings, coined QW on a line can be implemented. The actual direction of the walk is horizontal in Fig.~\ref{walkfig}, while the number of steps increases vertically. A network for QW of $M$ steps has $M+1$ outputs and it consists of $M(3M+5)/2$ elementary rings.

Let us note that the possible QW realization summarized above uses reflectionless rings as building blocks. The scheme does not involve optimization, it is based on analytic results obtained for single rings\cite{FMBP05a,FKBP05,KFBP08}. All types of rings can be realized with several parameter pairs (size and SOI strength), although these optimal parameters form discrete sets. Therefore the system can tolerate geometrical errors to some extent, but the optimization of a certain, not necessarily perfect geometry can be more difficult in this case than for rectangular networks.

\section{Stability: random scatterers}
When investigating the possible effects of scatterers, the most obvious one is the decrease of the conductance of the device. This consequence of the non-ideal transport is inevitable, uncontrollable scattering events involve backscattering as well, thus the devices themselves can not be made reflectionless. Therefore the decrease of the transmission probability (or the efficiency of the device) as a function of the strength of the scattering induced disturbances is the first question to investigate.

\begin{figure}[tbh]
\includegraphics[width=8cm]{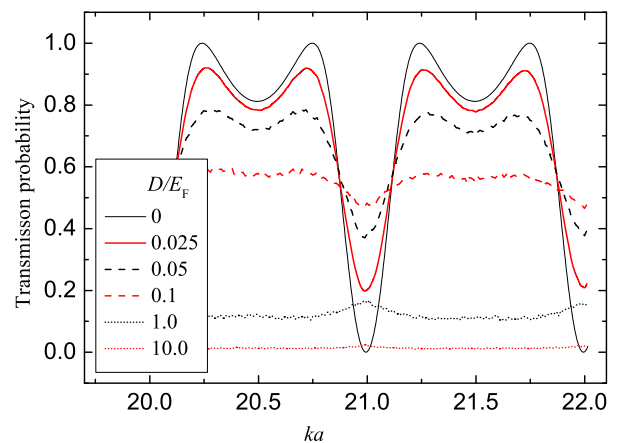}
\caption{(Color online) The transmission probability of a single quantum ring with diametrically coupled input and output leads for different strengths of the scattering events. The strength of the SOI is chosen such that $\omega/\Omega=1.0$ (see Eq.~(\protect\ref{Ham})).}
\label{SDfig}
\end{figure}

However, for some applications the most important point can be whether the electron's spatial and/or spin state is close to the desired one, provided they are transmitted at all. Reflective losses are less crucial in this case, we are mainly concerned in the transmitted wave function, and analyze to what extent it can be considered to be acceptable from the viewpoint of the functionality of the device.  This brings us to the question when does phase randomization (i.e., the second effect of scattering) become important?

Clearly, the two issues above are not independent, as one can check e.g.~for one dimensional propagation along a line, stronger dephasing means stronger reflection as well. In a network, however, there is an additional, less direct connection between (partial) reflections and effective dephasing. Namely, when the functionality of a certain device strongly relies on multiple coherent quantum mechanical interferences, reflections at internal points modify the wave functions that interfere at the junctions, leading to an additional, dephasing-like effect.

Fig.~\ref{SDfig} can serve as a reference, it shows for various values of $D$ the transmission probability of a single quantum ring as a function of the incoming wave number $k$ multiplied by the ring radius $a$. (The input and output leads are connected to this ring diametrically, and the strength of the $SOI$ is moderate: $\omega/\Omega=1.0$) When the ratio of $D$ (measured in energy units) and the Fermi energy is about $0.1,$ the transmission probability decreases from unity  below 60$\%.$  The contrast of the minimal and maximal transmission probabilities also tends to disappear when $D$ increases. For extremely strong scattering effects the transmission probability becomes negligible.

\bigskip

\begin{figure}[tbh]
\includegraphics[width=8cm]{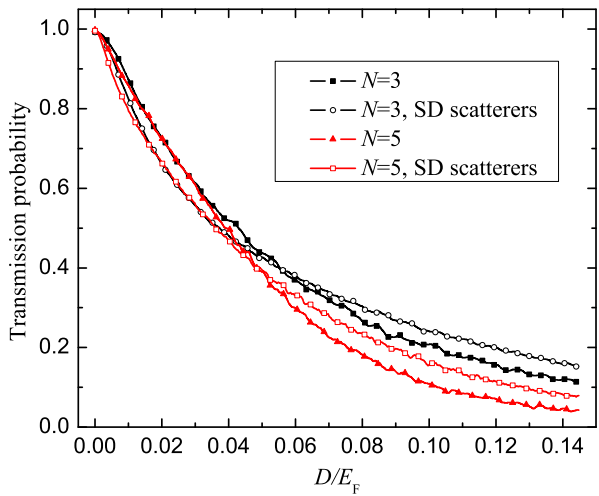}
\caption{(Color online) The transmission probability of rectangular $N\times N$ quantum ring arrays as a function of the strengths of the scattering events. The ring sizes and the SOI strengths correspond to the ideal operation described in the text.}
\label{NNtransfig}
\end{figure}

Now let us investigate rectangular arrays. The decrease of the transmission probability is shown by Fig.~\ref{NNtransfig} for both $3\times 3$ and $5 \times 5$ arrays. The parameters used for these plots are the optimal ones for the ideal case (without scatterers), that is, the intended operation of the smaller network is the direction of the input current to one of its output ports, while the $5 \times 5$ array should work analogously to the Stern-Gerlach device. As we can see, in the range of moderate values of $D$ (which is of our primary interest), spin-dependent (SD) scatterers cause the transmission probabilities to decrease faster than in the non-SD case. A weak size effect is also visible: the same density of scatterers cause slightly stronger dephasing in a larger network. It is important to mention here, that although there are considerable number of rings in these networks, the rate at which the transmission probabilities decrease as a function of $D,$ is not orders of magnitude higher than it is for a single ring.
As we mentioned in the introductory part of this section, increasing reflection probability is an inevitable consequence of the scattering events, but their presence does not necessarily lead to a complete destruction of the functionality of the device.
\begin{figure}[tbh]
\includegraphics[width=8cm]{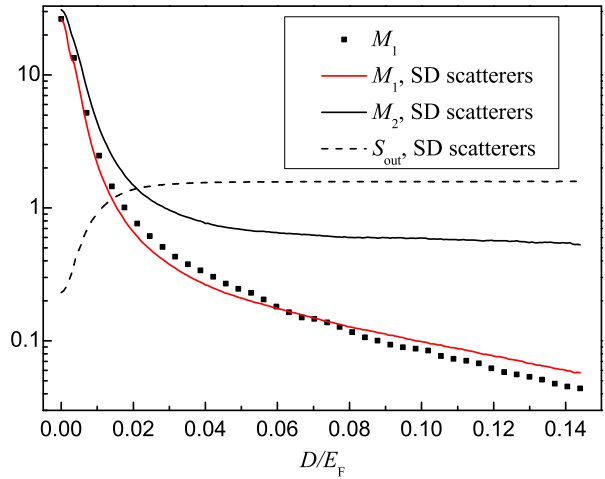}
\caption{(Color online) The measures of functionality given by Eq.~(\protect\ref{NNmeasures3}) for a rectangular $3\times 3$ quantum ring array as a function of the strengths of the scattering events. The ring sizes and the SOI strengths correspond to the ideal operation described in the text.}
\label{NNmfig3}
\end{figure}

In the case when a $3 \times 3$ rectangular network (see Fig.~\ref{NNfig}) is intended to direct its output current, various measures of the "distance" between its ideal and actual operation can be formulated in terms of the relevant probabilities:
\begin{equation}
\begin{aligned}
&M_1=\frac{P_1}{R+P_2+P_3}=\frac{P_1}{1-P_1}, \\  &M_2=\frac{P_1}{P_2+P_3}=\frac{P_1}{T-P_1}, \\ &S_{out}=-\sum_k \frac{P_k}{T}\log_2(\frac{P_k}{T}),
\label{NNmeasures3}
\end{aligned}
\end{equation}
where $R$ and $T$ denote the reflection and transmission probabilities, and $P_k$ stands for the probability that the electron leaves the device through output arm $k$ ($k=1$ corresponds to the distinguished output.) The measures $M_1$ and $M_2$ are signal to noise ratios, with $M_1$ being the more strict one in the sense that $M_2$ does not take reflective losses into account: $M_2$ compares the probability of the desired output to that of the unwanted ones.
$S_{out}$ is the Shannon entropy of the output, it should be zero in the ideal case, and has a maximum when all $P_k$ have the same values.  The dependence of these measures on the strength of the disturbances caused by the scatterers is shown in Fig.~\ref{NNmfig3}. As we can see, these measures decrease faster than the transmission probability, but acceptable signal to noise ratios can be seen even when the transmission probability is below 80\%. Note that the difference between effects caused by SD and non-SD scatterers is surprisingly small, they destroy quantum mechanical coherence at a similar rate. This is a general result for all the investigated devices, in the following we will focus on the spin-dependent case.
\begin{figure}[tbh]
\includegraphics[width=8cm]{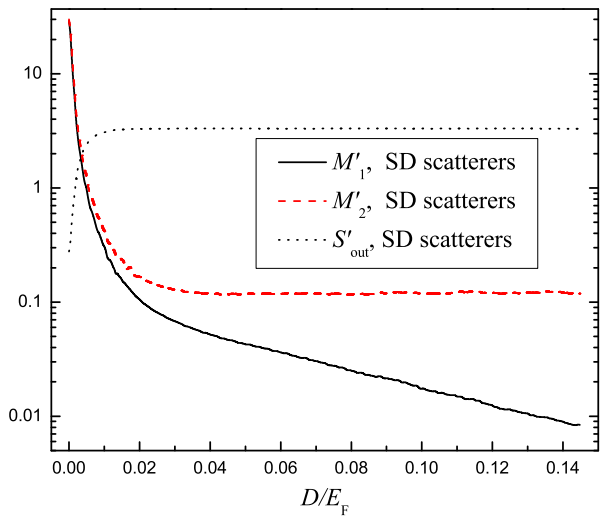}
\caption{(Color online) The measures of functionality given by Eq.~(\protect\ref{NNmeasures5}) for a rectangular $5\times 5$ quantum ring array as a function of the strengths of the scattering events. The ring sizes and the SOI strengths correspond to the ideal operation described in the text.}
\label{NNmfig5}
\end{figure}

Next we focus on the spintronic analogue of the Stern-Gerlach device proposed in Ref.~[\onlinecite{KFBP08c}]. According to the functionality of this device, the relevant measures are spin-dependent:
\begin{equation}
\begin{aligned}
&M_1^\prime=\frac{P_{\uparrow \uparrow 1}}{1-P_{\uparrow \uparrow 1}}, \\
&M_2^\prime=\frac{P_{\uparrow \uparrow 1}}{T-P_{\uparrow \uparrow 1}}, \\
&S_{out}^\prime=-\sum_k \frac{P_{\uparrow \uparrow k}}{T}\log_2(\frac{P_{\uparrow \uparrow k}}{T}) +\frac{P_{\uparrow \downarrow k}}{T}\log_2(\frac{P_{\uparrow \downarrow k}}{T}) \label{NNmeasures5}
\end{aligned}
\end{equation}
where we considered the case of spin-up input, and $P_{\uparrow \uparrow 1}=P_1(\mathrm{in:}\uparrow, \mathrm{out:}\uparrow)$ is the probability of the desired output.  $P_{\uparrow \downarrow k}=P_k(\mathrm{in:}\uparrow, \mathrm{out:}\downarrow)$ in the spin resolved Shannon entropy is related to spin-down output at the channel $k$ for spin-up input. (Note that similar measures can be defined also for the orthogonal input, but due to symmetry, those measures give the same numerical values as the ones above.)

Fig.~\ref{NNmfig5} shows the dependence of the measures given by Eq.~(\ref{NNmeasures5}) on $D.$ Note that the limiting value of $S_{out}^\prime$ for strong scattering events (and also that of $S_{out}$ shown in Fig.~\ref{NNmfig3}) is essentially the maximum of the Shannon entropy, i.e., in this limit the (weak) output current is distributed evenly among the output channels. Comparing Figs.~\ref{NNmfig5} and \ref{NNmfig3} it can be seen that the same value of $D$ induces considerably stronger effects in the larger network, although even the latter can be functional for a moderate level of disturbances. Size effect, however, can be analyzed best in QW networks, where the desired functionality is the same, it is only the number of the constituent rings that is different for networks that are designed to realize different number of QW steps.

\bigskip

In the case of the quantum walk, however, determining the spatial current distribution at the output (i.e, the probabilities corresponding to the transmission through the respective outputs) is not sufficient to quantify the effectiveness of the device: phase relations are also of fundamental importance. Therefore we calculate the overlaps
\begin{equation}
F_1=\mathrm{Tr}(\rho_{out} \rho_{i}),\  F_2=\frac{\mathrm{Tr}(\rho_{out} \rho_{i})}{T}, \ F_3=\frac{\mathrm{Tr}(\rho_{out} \rho_{i})}{T^\prime} \label{Wmeasures}
\end{equation}
where $\rho_{out}$ is given by Eq.~(\ref{rhoscatt}), and $\rho_{i}$ is the density operator corresponding to the pure quantum mechanical state that would result from the ideal QW described by Eq.~(\ref{QWstep}). In contrast to the transmission probability $T,$ that does not distinguish the outputs,  $T^\prime$ denotes the probability of transmission into "valid" output channels: In the ideal case there is no output into the leads at the sides of the device, the input current is distributed among the output leads located at the bottom side of Fig.~\ref{walkfig}. However, scattering effects modify this picture, and $T-T^\prime,$ the transmission probability into the side leads (that is zero in the ideal case), can also be considered as a measure of the functionality of the device.
\begin{figure}[tbh]
\includegraphics[width=8cm]{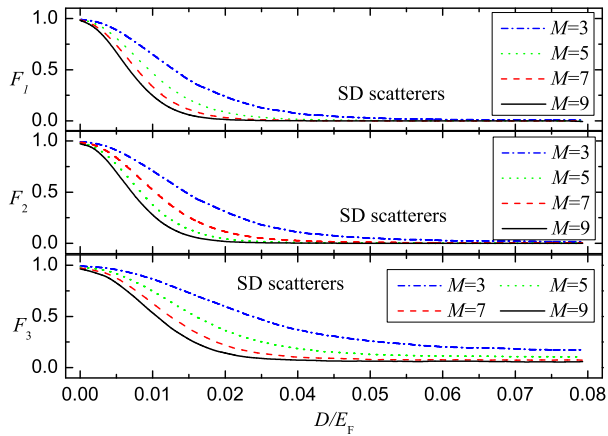}
\caption{(Color online) The measures of functionality given by Eq.~(\protect\ref{Wmeasures}) for quantum ring arrays to realize $M$ step QW. The ring sizes and the SOI strengths correspond to the ideal operation described in the text.}
\label{WDmfig}
\end{figure}

Considering transmission probabilities $T$ of networks that are designed to realize QW, the most remarkable point is that the functions $T(D)$ are quite similar for different network sizes (relative errors are below 5\% for $M=1,\ldots,9$ QW steps). If we recall that the input part of these devices of any size is exactly the same, the conclusion is that a small subnetwork close to the input junction is responsible for the reflection, i.e., scattering events in this small area determine the reflection probability $R=1-T.$ The measures $F_1, F_2$ and $F_3,$ however, show stronger size dependence, thus the current is distributed among the output channels in different ways for networks of different sizes. The overlaps given by Eq.~(\ref{Wmeasures}) are shown in Fig.~\ref{WDmfig} as a function of the strength of the scattering events. Recalling that the number of the elementary rings in a network to realize $M$ QW steps is proportional to $M^2,$ one might expect the overlaps $F_k$ measuring the fault tolerance of the devices to scale also as $M^2.$ However, choosing a certain level of any of these measures (e.g. $F_2=0.9$), we find that the $D$ values corresponding to this level do not decrease quadratically as the size of the networks increases. Instead, this decrease is even weaker than a linear dependence, thus the stability of the devices scale with size in a promising way. Note that similar error analysis for the possible quantum optical realizations of QW has been investigated in Refs.~[\onlinecite{KJ06,KBH06}].

\section{Stability: temperature related effects}
The role of finite temperatures are taken into account here as the appropriate broadening of the energy distribution of the incoming electrons. In view of this, there is a natural limit, describing the case when interference related effects are expected to appear at all. Namely, if there is a typical wave number $k,$ and a characteristic length $a$ in a quantum interference device, then interference properties are (quasi)periodic as a function of $ka.$ (In the case of a single ring with moderate SOI strength, $k$ can be the wave number of the input wave, which is assumed to be monoenergetic at this point, while $a$ is the radius of the ring.) Obviously, if the incoming wave numbers have a range of $\Delta k,$ $a\Delta k$ cannot be too large (i.e., comparable to unity), otherwise the interferences are smeared out. As we can see, there is a size effect here: at a given temperature $T$, the width of the equilibrium energy distribution is fixed, it has the order of $k_BT$, and $\Delta k$ is also given by the dispersion relation. Thus if we can decrease $a \Delta k $ e.g. by building smaller devices, then temperature related effects will become less pronounced. In other words, miniaturization, besides its obvious advantages, can help to avoid interferences to disappear (and also decrease dephasing caused by random scattering, simply by increasing the mean free path -- system size ratio).
\begin{figure}[tbh]
\includegraphics[width=8cm]{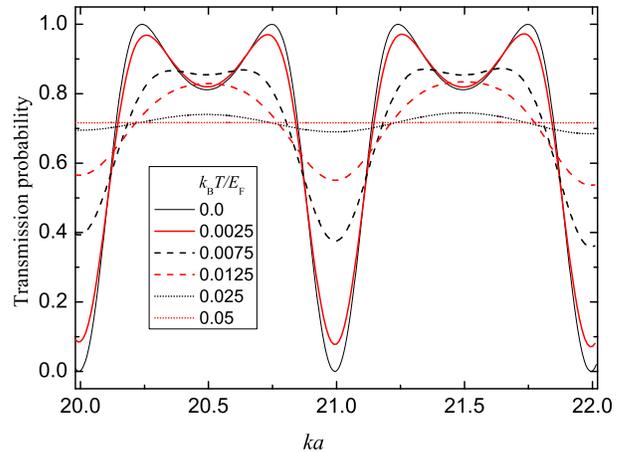}
\caption{(Color online) The transmission probability of a single quantum ring with diametrically coupled input and output leads for different temperatures. The strength of the SOI is characterized by $\omega/\Omega=1.0$ (see Eq.~(\protect\ref{Ham})).}
\label{STfig}
\end{figure}

Before we analyze more complex systems, let us consider again a single quantum ring with one input and one output arm, with $ka$ being in the experimentally achievable range around 20. Fig.~\ref{STfig} shows -- in accordance with the introductory part of this section -- that the transmission probabilities are noticeably modified already when $k_B T$ is a small portion of the Fermi energy. (Numerically, for $E_F$ being in the range of 10 meV, the flat line that belongs to the highest temperature in Fig.~\ref{STfig}, is still in the cryogenic range.) Comparing the two limits when non-monoenergetic input and strong scattering events completely change the transmission probabilities (see Figs.~\ref{STfig} and \ref{SDfig}), an important difference becomes apparent: high temperatures lead to finite transmission probabilities, while strong scattering events cause the device to be opaque. The latter result can be understood readily, but the first one may need some explanation. High temperature conductance is an integral property in the sense of Eq.~(\ref{rhoT}), if the input is normalized, we may write\cite{D95}
\begin{equation}
G(T)=\frac{2 e^2}{\hbar}\int p(E,T)
\left\langle\Psi_{out}(E) |\Psi_{out}(E)\right\rangle dE.
\end{equation}
Thus all monoenergetic transmission probabilities contribute to the conductance with a weight determined by the energy distribution. Recalling that the transmission probability is a quasiperiodic function of $ka,$ for wide enough input energy distribution (being almost constant during a period) the averaged conductance is proportional to the integral of the transmission probability over a period divided by the length of the period. For a given geometry, this quantity is a constant, but it can change when the strength of the SOI is modified.

\begin{figure}[tbh]
\includegraphics[width=8cm]{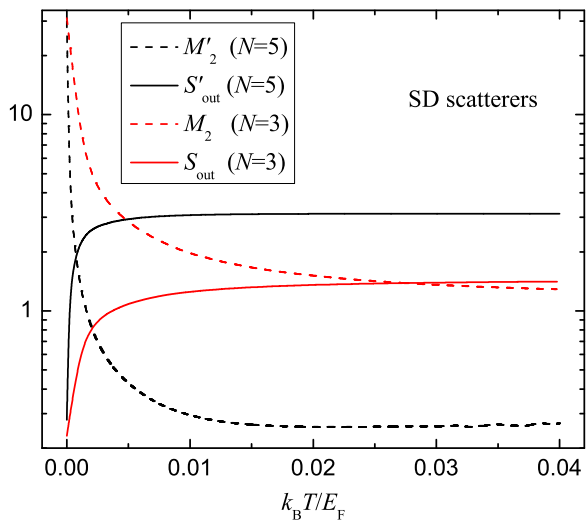}
\caption{(Color online) The measures of functionality given by Eqs.~(\protect\ref{NNmeasures3}) and (\protect\ref{NNmeasures5}) for rectangular $3\times 3$ and $5\times 5$ quantum ring arrays as a function of temperature. The ring sizes and the SOI strengths correspond to the ideal operation described in the text.}
\label{NNTmfig}
\end{figure}

\begin{figure}[tbh]
\includegraphics[width=8cm]{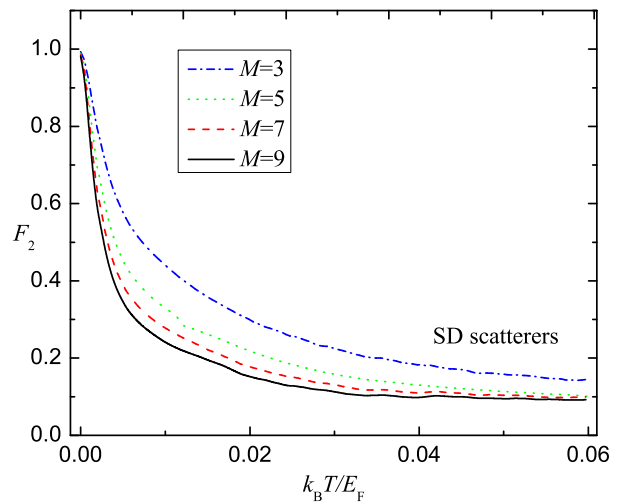}
\caption{(Color online) The measures of functionality given by Eq.~(\protect\ref{Wmeasures}) for quantum ring arrays to realize $M$ step QW. The ring sizes and the SOI strengths correspond to the ideal operation described in the text.}
\label{WTmfig}
\end{figure}

Fig.~\ref{NNTmfig} shows that a $3\times 3$ rectangular array can direct its input current to one of its output ports with $M_2=2$ signal to noise ratio at a temperature, where the transmission probability of a single ring reduces already below $80\%$ of its initial value (see Fig.~\ref{STfig} as well.) The spintronic analogue of the Stern-Gerlach device based on a $5\times 5$ network is more sensitive to thermal fluctuations, which is due to its larger size and different functionality. (It is worth noting that the Shannon entropies reach their maxima in the high temperature limit as well.)  According to Fig.~\ref{WTmfig}, the stability of the QW networks scale with their size $M$ similarly to the case of random scatterers: given values of $F_2$ is reached at temperatures that decrease as a function of $M$ slower than linearly.

\bigskip

These results, however, still mean that quantum interference devices require quasi-monoenergetic input electrons, otherwise interference effects are smeared out. In usual semiconductors, the limit temperature is quite low, meaning an obstacle in many practical applications. This effect can be decreased by building smaller devices, or decreasing the width of the input energy distribution below the thermal equilibrium. Note that in this latter case nonelastic scattering events will eventually lead to thermal equilibrium again, but the thermalization length being relevant here can be considerably larger than e.g.~the dephasing length.

As a final point, we investigated whether simple network-based devices can provide sufficiently narrow transmission peaks, i.e., whether they can deliver quasi-monoenergetic electrons. (We note that this might not be the most effective method to overcome the problem caused by thermal broadening of the input energies, but it can serve demonstrative purposes.) We found that an ideal linear chain consisting of five rings of linearly increasing sizes ($+0.5\%$ for the consecutive rings) has a transmission peak the width of which corresponds to $k_BT=10^{-5}\times E_F.$ The transmission probability is essentially zero in an energy interval of $k_B/E_F=0.06$ (At the peak it is around $60\%.$) These ideal properties are of course modified with introducing scatterers like we did in the current paper. However, it was found that for moderate strength of the scattering effects (e.g.~when the overlap $F_3$ is around 0.5 for a nine step QW), the height of the transmission peak is still above $10\%$, and its characteristic width corresponds to $k_BT=10^{-4}\times E_F$, which is remarkable as we did not optimize the geometry for these values.

\section{Summary}
In this paper we investigated the stability of network-based spintronic devices against random scattering events and thermal fluctuations. We focused on the question to what extent the functionality of certain proposed networks remains close to the ideal behavior when temperature or the intensity of the scattering induced disturbance increases. We considered two different network types, a rectangular array that can serve as a spintronic analogue of the Stern-Gerlach apparatus as well as networks that can realize quantum walk with the spin degree of freedom playing the role of the quantum coin. We introduced appropriate measures to quantify the "distance" between the ideal and the realistic behavior of these devices, and found that the functionality can tolerate moderate level of errors. It was demonstrated that although networks of increasing size are less stable, but fortunately this decrease of stability is slow with increasing the size of the networks.

\section*{Acknowledgments}

We thank M.~G.~Benedict and T.~Kiss for the useful discussions. This work was supported by the
Flemish Science Foundation (FWO-Vl), the Belgian Science Policy (IAP) and the
Hungarian Scientific Research Fund (OTKA) under Contracts Nos.~T48888,
M36803, M045596. P.F.~was supported by a J.~Bolyai grant of the Hungarian
Academy of Sciences.

\end{document}